\documentclass[letterpaper]{article}

\usepackage{amsmath,amssymb}
\usepackage{subfigure}
\usepackage{tikz}
\usepackage{algorithm}
\usepackage[margin=1.1in]{geometry}
\usetikzlibrary{trees}

\usepackage{url}

\newcounter{classcounter}
\newtheorem{class}[classcounter]{Class}

\newcommand{\G}{\mathcal{G}}
\newcommand{\J}{\mathcal{J}}
\newcommand{\T}{\mathcal{T}}

\newcommand{\I}{\mathcal{I}}
\newcommand{\SG}{\mathcal{S}_{\G}}
\newcommand{\ST}{\mathcal{S}_{\T}}

\newcommand{\AD}{App}
\newcommand{\DD}{Dis}
\newcommand{\temp}[1]{\hat{#1}}
\newcommand{\topo}[1]{{#1}}
\newcommand{\ecc}{\varepsilon}
\newcommand{\TVG}{\ensuremath{\G=(V,E,\T,\rho,\zeta)}}

\newcommand{\ct}[1]{\textit{(#1)}:}

\title{Time-Varying Graphs and Dynamic Networks\thanks{A preliminary version of this paper appeared in proceedings of the 10th International Conference on Adhoc Networks and Wireless ({\sc Adhoc-Now'11)}.}}

\author{Arnaud Casteigts$^{\dagger}$, Paola Flocchini$^{\dagger}$, Walter Quattrociocchi$^{\ddagger}$, Nicola Santoro$^{\S}$\bigskip\\
\small
$^\dagger$University of Ottawa, Canada\vspace{-3pt}\\
\footnotesize
{\tt \{casteig,flocchin\}@site.uottawa.ca}\smallskip\\
\small
$^\ddagger$University of Siena, Italy\vspace{-3pt}\\
\footnotesize
{\tt walter.quattrociocchi@unisi.it}\smallskip\\
\small
$^\S$Carleton University, Ottawa, Canada\vspace{-3pt}\\
\footnotesize
{\tt santoro@scs.carleton.ca}\medskip\\
}
\date{}

\begin{document}

\maketitle

\begin{abstract}
The past few years have seen intensive research efforts carried out in some apparently unrelated areas of dynamic systems --~delay-tolerant networks, opportunistic-mobility networks, social networks~-- obtaining closely related insights. Indeed, the concepts discovered in these investigations can be viewed as parts of the same conceptual universe; and the formal models proposed so far to express some specific concepts are components of a larger formal description of this universe. The main contribution of this paper is to integrate the vast collection of concepts, formalisms, and results found in the literature into a unified framework, which we call {\em TVG} (for {\em time-varying graphs}). Using this framework, it is possible to express directly in the same formalism not only the concepts common to all those different areas, but also those specific to each. Based on this definitional work, employing both existing results and original observations, we present a hierarchical  classification of TVGs; each class corresponds to a significant property examined in the  distributed computing literature. We then examine how TVGs can be used to study the evolution of network properties, and propose different techniques, depending on whether the indicators for these properties are a-temporal (as in the majority of existing studies) or temporal. Finally, we briefly discuss the introduction of randomness in TVGs.
\end{abstract}~\\

{\footnotesize \noindent
\textbf{keywords:}
Delay-tolerant networks; opportunistic networks; social networks; dynamic graphs; time-varying graphs; distributed computing.}

\section{Introduction}

In the past few years, intensive research efforts have been devoted to some
apparently unrelated areas of dynamic systems,
obtaining closely related insights. This is  particularly evident in  (a)
the study of communication  in highly dynamic networks,
e.g.,  broadcasting and routing in  delay-tolerant networks;
(b) the exploitation of  passive mobility, e.g., the opportunistic use
of transportation networks; and
(c) the analysis of complex real-world networks ranging from neuroscience or biology to transportation systems or social studies,
e.g., the characterization of the interaction patterns
emerging in a social network.

As part of these research efforts, a number of important
concepts have been identified, often named, sometimes formally defined.
Interestingly, it is becoming apparent  that  these concepts are
strongly related. In fact, in several cases,
differently named  concepts identified by different researchers are
actually one and  the same concept.  For example,
the concept of {\em temporal distance},  formalized in~\cite{BFJ03},  is the same as {\em reachability time}~\cite{Holme05}, {\em information latency}~\cite{KosKW08}, and  {\em temporal proximity}~\cite{Kostakos09};
 similarly, the concept of {\em journey} \cite{BFJ03} has been called {\em schedule-conforming path}~\cite{Berman96}, {\em time-respecting path}~\cite{Holme05,KKK00}, and {\em temporal path}~\cite{ChMMD08,TSM+09}.
Hence,  the notions discovered in these investigations  can be
viewed as parts of the same conceptual universe; and the formalisms proposed so far to express some specific
concepts can be viewed as fragments of a larger formal description
of this universe.
A common point in all these areas is that the system structure - {\em the network topology} -  varies in time.
Furthermore the rate and/or degree of the changes is generally too high to be reasonably modeled in terms of network faults or failures: in these systems
{\em changes are not anomalies but rather integral  part of the nature of the system}.

As the notion of  (static) graph is the natural means for representing a static network,
the notion of dynamic (or {\em time-varying}, or {\em evolving}) graph  is the natural means
to represents these highly dynamic networks. All the concepts and  definitions  advanced so far are based  on or imply such a notion, as expressed  even by the choices of names; e.g.,
Kempe et al. \cite{KKK00} talk of a {\em temporal network} $(G,\lambda)$ where $\lambda$ is a {\em time-labeling} of the edges, that associates {\em punctual} dates to represent dated interactions; Leskovec et al. \cite{LeKF07}  talk of {\em graphs over time};
Ferreira ~\cite{Fer04} views the dynamic of the system in terms of
 a sequence of static graphs, called an {\em evolving graph};   Flocchini et al. \cite{FMS09}
 and Tang et al.~\cite{TMML10} independently employ the term {\em time-varying graphs}; Kostakos uses the term {\em temporal graph} \cite{Kostakos09};  etc.

The main contribution of this paper is to
integrate the existing models, concepts, and results proposed in the literature into a unified  framework,
which we call {\em TVG}  (for   {\em time-varying graphs}).
Using it, it is possible to express directly
in the same formalism not only the concepts common to
all these different areas, but also those specific to each.
This, in turns, should enable the transfer of results from one application
area to another.

The paper first provides background motivation in Section~\ref{sec:contexts}, by mentioning a range of works where the need for dynamics-related concepts emerged. Section~\ref{sec:TVG} presents the TVG formalism together with dedicated notations. This formalism is used and extended in Section~\ref{sec:concepts}, where we present the most central concepts that have been identified by the research ({\it e.g.},  journey, temporal subgraphs, distance and connectivity); we also address the
different  perspective ({\it e.g.,} the graph-centric  (or global) point of view  {\it vs.}
the edge-centric (or interaction based) point of view).
The paper then continues into two main blocks.

The first block, Section~\ref{sec:classes}, more oriented towards the field of distributed computing, examines
the impact of properties of TVGs on the feasibility and complexity of distributed problems, reviewing and unifying a large body of literature. In particular, we identify several classes of TVGs  defined with respects to basic properties on the network dynamics. Some of these classes have been extensively studied in different contexts; e.g., one of the TVG classes considered here coincides with the family of dynamic graphs over which {\em population protocols} (\cite{AAD+06,AAER07}) are defined. We examine the (strict) inclusion  hierarchy among the classes. To several of the class-defining properties  considered here correspond  necessary conditions and impossibility results for basic computations. Thus, the inclusion relationship implies that we can transfer feasibility results (e.g., protocols) to an included class, and impossibility results (e.g., lower bounds) to an including class.

The second block in Section~\ref{sec:analysis} is concerned with dynamic network analysis. We deal with three aspects in particular: the automated verification of deterministic properties on network traces; how temporal concepts can be leveraged to express new phenomenon or properties in complex systems; and how TVGs could be used to study a coarser-grain evolution of network properties. Different techniques are proposed for the latter, depending on whether the indicators for these properties are {\em a-temporal} (as in the majority of existing studies) or {\em temporal}, that is, based on properties that take place over time such as the concept of journey, temporal distance and connectivity.

Finally, in Section~\ref{sec:stochastic}   we discuss the introduction of randomness in TVGs, and review results.

In addition to the new results and perspectives, and besides the {\em de facto} survey that these sections represent, the main contribution of this paper certainly remains that of integrating all the reviewed material within a single and unified formalism.

\section{Contexts}
\label{sec:contexts}

We mention below three research areas in which dynamical aspects have played a central role recently. They include {\em delay-tolerant networks}, {\em opportunistic-mobility networks}, and {\em real-world complex networks}. Interestingly, these areas have seen a number of similar concepts emerge with distinct purposes, ranging from the design of solutions in delay-tolerant networks to the analysis of phenomena in complex dynamic network.

\subsection{Delay-Tolerant Networks}

Delay-tolerant networks are highly-dynamic, infrastructure-less networks whose essential characteristic is a possible absence of end-to-end communication routes at any instant. These networks, also called {\em disruptive-tolerant},  {\em challenged},  or {\em opportunistic}, include for instance satellite, pedestrian, and vehicular networks. Although the assumption of {\em connectivity} does not necessarily hold at a given instant -- the network could even be disconnected at every time instant -- communication routes are generally available over time and space, enabling for example broadcast and routing by means of a store-carry-forward-like mechanism.

An extensive amount of research has been recently devoted to these types of problems ({\it e.g.} \cite{BGJL06,CLW07,JMR10,JFP04,LW09b,MSG08,RDBT12,SPR05,Zha06}). 
A number of new routing and broadcast techniques were designed to face such an extreme context, based for example on pro-active knowledge on the network schedule~\cite{JFP04, BFJ03}, probabilistic strategies~\cite{LDS03, SPR05}, delay-based optimization~\cite{RRS11}, or encounter-based choices~\cite{GroV03,JLW07}. 
Other recent works considered the broadcast problem from an analytical and probabilistic standpoint, e.g., in~\cite{BCF09,CleMMPS08} where the maximal propagation speed is characterized as a function of the rate of topological changes in the network (these changes are themselves regulated by Markovian processes on edges). 
In all these investigations, the time dimension has had a strong impact on the research, and led the research community to extend most usual graph concepts -- e.g, paths and reachability~\cite{Berman96,KKK00}, distance~\cite{BFJ03}, diameter~\cite{ChMMD08}, or connected components~\cite{BF03} -- to a temporal version.

\subsection{Opportunistic-Mobility Networks}

As mobile carriers and devices become increasingly equipped with short-range radio capabilities,  it is   possible to 
 exploit the (delay-tolerant) networks created by their mobility for uses that are possibly external and extraneous to the carriers.
In fact, other  entities  (e.g., code, information, web pages) called {\em agents}  can  opportunistically   ``move"  on the carriers'  network for their own purposes,
by using the mobility of the carriers (sometimes called {\em ferries}) as a transport mechanism.
 Such networks have been deployed e.g, in the context
of buses \cite{BZCLV07, BGJL06},  and pedestrians  \cite{ChHCDGS07}.
 Example of carrier networks and opportunistic mobility usages  include:  {\em Cabernet},
  currently deployed in 10 
taxis running in the Boston area \cite{EBM08}, 
which
 allows to deliver messages 
and files 
to users in cars;
 and 
  {\em UMass DieselNet},  consisting of WiFi nodes attached to 
40 buses in Amherst, used for routing, information delivery,
and connectivity measurements \cite{BGJL06,ZKL+07}.

Of particular interest is the class of carriers/ferries 
 following a deterministic periodic trajectory.
 This class naturally  includes infrastructure-less networks where mobile entities have fixed routes 
that they traverse regularly. Examples of such common settings are public transports, low earth orbiting (LEO) satellite systems, security guards' tours, etc.
These networks have been investigated with respect to routing and to the design of carriers' routes
 (e.g., see~\cite{GK07,LW09b}) and more specifically for buses (\cite{BZCLV07,ZKL+07}),
  and satellites~\cite{WTSB09}. 
In addition to routing,  some algorithmic works have been done in the contexts of network exploration~\cite{FMS09, IW11, FKMS12} and creation of broadcast structures~\cite{CasFMS10}. 
In the derivation of these results, the temporal component has played a crucial role, both in terms of extension of concepts and of developing solution techniques.

\subsection{Real-World Complex Networks}

The research area of {\em complex systems} addresses the analysis of real complex dynamic networks, ranging from neuroscience and biology to transportation networks and social studies, with a particular interest in the understanding of self-organisation, emergence properties, and their reification.

As stated in    \cite{LESK10}, the central problem in this area is the definition of mathematical models able to capture and to reproduce properties observed on the real dynamics of the networks (e.g., shrinking diameter~\cite{LeKF07}, formation of communities, or appearance of inequalities). 
A fundamental work  on graphs where edges are endowed with temporal properties is the one by Kempe and Kleinberg  \cite{KeK02}, in which  the basic properties (both combinatorial and algorithmic) of graphs are addressed when the connections among nodes are constrained by temporal conditions. The formalism introduced therein to represent dynamic graphs has been used as framework for several works such as~\cite{BHKL06, Eagle06, kempe03, Scherrer08}.

In \cite{Kostakos09} the theoretical framework of temporal graphs is proposed  to study a large dataset of emails records. The author proposed to label graphs with temporal attributes by allowing the representation of each node as a chain of all its temporal instances during time;
some interesting metrics aimed at capturing the interactions among nodes during time, e.g., temporal or geodesic proximity, are discussed. 
In \cite{TSM+09}  an extension  of the  model of  \cite{KeK02}  is proposed 
by looking at the smallest delay path in generic information spreading process. 
The authors try to overcome the limits of the previous works (mainly concerned with local aspects) by defining a temporal graph as a sequence of static graphs whose elements aggregate all interactions during given time-windows -- we will call such construct a {\em sequence of footprints}.
In \cite{KosKW08} the authors study the temporal dynamics of communication over a dataset of on-line communications and emails over a two years period.
The main metric introduced to capture the interaction is again the temporal distance, defined there as the minimum time needed for a piece of information to spread from an individual to another by means of multihop sequences of emails.

As these investigations indicate, temporal concerns are an integral part of recent research efforts in complex systems. It is also apparent that the emerging concepts are in essence the same as those from the field of communication networks, involving again temporal definitions of the notions of paths, distance, and connectivity, as well as many higher concepts that we identify in this paper.

\section{Time-Varying Graphs}
\label{sec:TVG}

Consider a set of entities $V$ (or {\em nodes}), a set of relations $E$ between these entities ({\em edges}), and an alphabet $L$ accounting for any property such a relation could have ({\em label}); that is, $E \subseteq V \times V \times L$. The definition of $L$ is domain-specific, and therefore left open --~a label could represent for instance the intensity of relation in a social network, a type of carrier in a transportation network, or a particular medium in a communication network; in some contexts, $L$ could be empty (and thus possibly omitted). For generality, we assume $L$ to possibly contain multi-valued elements (e.g. $<${\it satellite link; bandwidth of 4\,MHz; encryption available;...}$>$). The set $E$ enables multiple relations between a pair of entities, as long as these relations have a distinct label.

Because we address dynamical systems, the relations between entities are assumed to take place over a time span $\T \subseteq \mathbb{T}$ called the {\em lifetime} of the system. The temporal domain $\mathbb{T}$ is generally assumed to be $\mathbb{N}$ for discrete-time systems or $\mathbb{R}^+$ for continuous-time systems. The dynamics of the system can be subsequently described by a time-varying graph, or TVG, \TVG, where

\begin{itemize}
\item $\rho: E \times \T \rightarrow \{0,1\}$, called {\em presence} function, indicates whether a given edge is available at a given time.
\item $\zeta: E \times \T \rightarrow \mathbb{T}$, called {\em latency} function, indicates the time it takes to cross a given edge if starting at a given date (the latency of an edge could vary in time).
\end{itemize}

The model can be naturally extended by adding a {\em node presence function} $\psi: V \times \T \rightarrow \{0,1\}$ ({\it i.e.,} the presence of a node is conditional upon time) and a {\em node latency function} $\varphi: V \times \T \rightarrow \mathbb{T}$ (accounting {\it e.g.} for local processing times). 

\begin{figure}[h]
  \centering
  \begin{tabular}{c|c}
    \subfigure[Transportation network]{
      \label{fig:example1}
      \begin{tikzpicture}[yscale=1.4, xscale=1.5]
        \tikzstyle{every node}=[draw, rectangle, rounded corners=4pt, font=\scriptsize]
        \path (0,0) node (o) {Ottawa};
        \path (1,1) node (m) {Montreal};
        \path (2,0) node (l) {Lisbon};
        \tikzstyle{every path}=[->]
        \tikzstyle{every node}=[font=\scriptsize]
        \draw (o) edge[bend left=60] node[left] {$\lambda_1$} (m);
        \draw (o) edge[bend left=40] node[right] {$\lambda_2$} (m);
        \draw (m) edge[bend left=40] node[left] {$\lambda_3$} (l);
        \draw (m) edge[bend left=60] node[right] {$\lambda_4$} (l);
        \path (o)+(0,-.3) coordinate;
      \end{tikzpicture}
    }
    &
    \subfigure[Communication network]{
      \label{fig:example2}
      \begin{tikzpicture}[scale=1.4]
        \tikzstyle{every node}=[draw, circle, minimum size=11pt, inner sep=0pt]
        \path (0,0) node (a){$a$};
        \path (a)+(0, 1) node (b){$b$};
        \path (a)+(1,.5) node (c){$c$};
        \path (c)+(1.2,0) node (d){$d$};
        \tikzstyle{every node}=[font=\scriptsize, inner sep=1pt]
        \draw (a)--node[midway, left]{$\lambda_1$}(b);
        \draw (a)--node[midway, xshift=-4pt, yshift=-1pt, below right]{$\lambda_1$}(c);
        \draw (b)--node[midway, xshift=-4pt, yshift=1pt, above right]{$\lambda_1$}(c);
        \draw (c) edge node[midway, below]{$\lambda_2$}(d);
        \path (a)+(0,-.3) coordinate;
      \end{tikzpicture}
    }
  \end{tabular}
  \caption{\label{fig:examples}Two examples of time-varying graphs, employed in very different contexts.}
\end{figure}
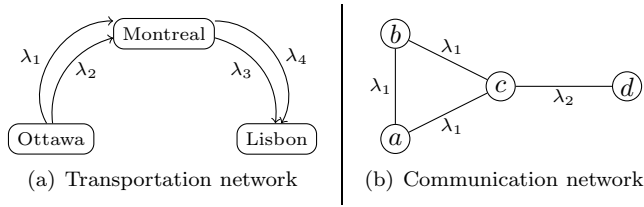

The TVG formalism can arguably describe a multitude of different scenarios, from transportation networks to communication networks, complex systems, or social networks. Two intuitive examples are shown on Figure~\ref{fig:examples}.

The meaning of what is an edge in these two examples varies drastically. In Figure~\ref{fig:example1}, an edge from a node $u$ to another node $v$ represents the possibility for some agent to {\em move} from $u$ to $v$. The edges in this example are assumed directed, and possibly multiple. The meaning of the labels $\lambda_1$ to $\lambda_4$ could be for instance ``{\it bus}'', ``{\it car}'', ``{\it plane}'', ``{\it boat}'', respectively. Except for the travel in car from Ottawa to Montreal --~which could assumably be started anytime~--, typical edges in this scenario are available on a {\em punctual} basis, {\it i.e.,} the presence function $\rho$ for these edges returns $1$ only at particular date(s) when the trip can be started. The latency function $\zeta$ may also vary from one edge to another, as well as for different availability dates of a same given edge (e.g. variable traffic on the road, depending on the departure time).

The second example on Figure~\ref{fig:example2} represents a history of connectivity between a set of moving nodes, where the possibilities of communication appear e.g. as a function of their respective distance. The two labels $\lambda_1$ and $\lambda_2$ may account here for different types of communication media, such as WiFi and Satellite, having various properties in terms of range, bandwidth, latency, or energy consumption. In this scenario, the edges are assumed to be undirected and  there is no more than one edge between any two nodes. The meaning of an edge is also different here: an edge between two nodes means that any one (or both) of them can (attempt to) send a message to the other. A typical presence function for this type of edge returns $1$ for some {\em intervals} of time, because the nodes are generally in range for a non-punctual period of time. Note that the effective delivery of a message sent at time $t$ on an edge $e$ could be subjected to further constraints regarding the latency function, such as the condition that $\rho(e)$ returns $1$ for the whole interval $[t, t+\zeta(e,t))$.

\begin{figure}[h]
  \centering
  \begin{tikzpicture}[scale=1.4]
    \tikzstyle{every node}=[draw, circle, minimum size=11pt, inner sep=0pt]
    \path (0,0) node (a){$a$};
    \path (a)+(0, 1) node (b){$b$};
    \path (a)+(1.1,.5) node (c){$c$};
    \path (c)+(1.4,0) node (d){$d$};
    \tikzstyle{every node}=[font=\scriptsize, inner sep=1pt]
    \draw (a)--node[midway, left]{$[1,3)$}(b);
    \draw (a)--node[midway, xshift=-4pt, yshift=-1pt, below right]{$[2,5)$}(c);
    \draw (b)--node[midway, xshift=-4pt, yshift=1pt, above right]{$[0,4)$}(c);
    \draw (c)--node[midway, above]{$[5,6)\cup [7,8)$}(d);
  \end{tikzpicture}
  \caption{\label{fig:simple-TVG} A simple TVG. {\it The interval(s) on each edge $e$ represents the periods of time when it is available, that is, $\cup(t\in \T : \rho(e,t) = 1)$.}}
\end{figure}
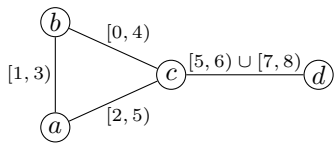

These two examples are taken different on purpose; they illustrate the spectrum of {\em models} over which the TVG {\em formalism} can stretch. As observed, some contexts are intrinsically simpler than others and call for restrictions (e.g.  between any two nodes  in the second example,  there is at most one 
undirected edge).
Further restrictions may be considered. For example the latency function could be decided constant over time ($\zeta: E \rightarrow \mathbb{T}$); over the edges ($\zeta: \T \rightarrow \mathbb{T}$); over both ($\zeta \in \mathbb{T}$), or simply ignored. In the latter case, a TVG could have its relations fully described by a graphical representation like that of Figure~\ref{fig:simple-TVG}.

A number of analytical work on dynamic networks simply ignore $\zeta$, or assume a discrete-time scenario where every time step implicitly corresponds to a constant $\zeta$. This value is also neglected in general when the graph represents dated interactions over a social network (the edges in this context are generally assumed to be punctual both in terms of instantaneous presence and null latency). The definitions we give in this paper address the general case, where \TVG.

\section{Definitions of TVG concepts}
\label{sec:concepts}

This section transposes and generalizes a number of dynamic network concepts into the framework of time-varying graphs. A majority of them emerged independently in various areas of scientific literature; some appeared more specifically; some others are original propositions.

\subsection{The underlying graph G}
\label{sec:underlying-graph}

Given a TVG \TVG, the graph $G=(V,E)$ is called {\em underlying} graph of $\G$. This static graph should be seen as a sort of {\em footprint} of $\G$, which flattens the time dimension and indicates only the pairs of nodes that have relations at some time in $\T$. It is a central concept that is used recurrently in the following. 

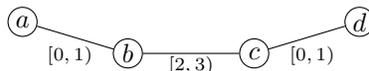
\begin{figure}[h]
  \centering
  \begin{tikzpicture}[scale=1.4]
    \tikzstyle{every node}=[draw, circle, minimum size=11pt, inner sep=0pt]
    \path (0,.3) node (a){$a$};
    \path (1, 0) node (b){$b$};
    \path (2.2,0) node (c){$c$};
    \path (3.2,.3) node (d){$d$};
    \tikzstyle{every node}=[font=\scriptsize, inner sep=1pt]
    \draw (a)--node[midway, below, xshift=-2pt, yshift=-2pt]{$[0,1)$}(b);
    \draw (b)--node[midway, below]{$[2,3)$}(c);
    \draw (c)--node[midway, below, xshift=2pt, yshift=-2pt]{$[0,1)$}(d);
  \end{tikzpicture}
  \caption{\label{fig:connected-G} An example of TVG that is not ``connected over time'', although its underlying graph $G$ is connected. {\it Here, the nodes $a$ and $d$ have no mean to reach each other through a chain of interaction.}}
\end{figure}

In most studies and applications, $G$ is assumed to be connected; in general, this is not necessarily the case.
Let us stress that  the connectivity   of $G=(V,E)$ does not imply that $\G$ is connected at a given time instant; in fact, $\G$ could be disconnected at all times. The lack of relationship, with regards to connectivity, between $\G$ and its footprint $G$ is even stronger:
 the fact that  $G=(V,E)$ is connected does not even imply that $\G$ is ``connected over time'', as illustrated on Figure~\ref{fig:connected-G}.

\subsection{Point of views}
Depending on the problem under consideration, it may be convenient to look at the evolution of the system from the point of view of a given relation (edge), a given entity (node), or from that of the global system (entire graph). We respectively qualify these views as \emph{edge-centric}, {\em vertex-centric}, and \emph{graph-centric}.

\subsubsection{Edge-centric evolution}
From an edge standpoint, the notion of evolution comes down to a variation of availability and latency over time. We define the {\em available dates} of an edge $e$, noted $\I(e)$, as the union of all dates at which the edge is available, that is, $\I(e)= \{t \in \T : \rho(e,t)=1\}$. 
When  $\I(e)$ is expressed as a multi-interval of availability $\I(e)=\{[t_1,t_2)\cup[t_3,t_4)...\}$, where $t_i<t_{i+1}$, the sequence of dates $t_1,t_3,...$ is called {\em appearance dates} of $e$, noted $\AD(e)$, and the sequence of dates $t_2, t_4,...$ is called {\em disappearance dates} of $e$, noted $\DD(e)$. Finally, the sequence $t_1, t_2, t_3,...$ is called {\em characteristic dates} of $e$, noted $\ST(e)$. In the following, we use the notation $\rho_{[t, t')}(e)=1$ to indicate that  $\forall t'' \in [t,t'), \rho(e,t'')=1$.

\subsubsection{Vertex-centric evolution}

From a node standpoint, the evolution of the network materializes as a succession of changes among its neighborhood.  This point of view does not appear frequently in the literature; yet, it was used for example in~\cite{OW05} to express dynamic properties in terms of   local variation of the {\em sequence of neighborhoods}
 $N_{t_1}(v), N_{t_2}(v)$.. where $N_{t}(v)$ denotes the neighbors of $v$ at time $t$ and  each $t_i$ corresponds to a date of local change (i.e., appearance/disappearance of an incident edge).

The {\em degree} of a node $u$ can be defined both in punctual or integral terms, e.g. with 
$Deg_t(u)=|E_t(u)|$, or $Deg_{\T}(u)=|\cup\{E_{t}(u) : t \in \T\}|$ where $E_t(u)$ indicates the set of edges incident
on $u$ at time $t$.

\subsubsection{Graph-centric evolution}

The sequence $\ST(\G) = sort(\cup\{\ST(e): e \in E\})$, called  {\em characteristic dates} of $\G$, 
corresponds to the sequence of dates when topological events 
(appearance/disappearance of an edge) occur in the system. Each topological event can be viewed
as the transformation
from one static graph to another.
Hence,  the evolution of the system  can be described as a sequence of
static graphs. More precisely,  from a global viewpoint,
the evolution of $\G$ is described as the sequence of graphs $\SG=G_{1}, G_{2}, ...$ where 
$G_i$  corresponds to the static {\em snapshot} of $\G$ at time $t_i\in \ST(\G)$;
i.e.,  $e\in E_{G_i} \iff \rho_{[t_i,t_{i+1})}(e)=1$.
Note that, by definition, $G_{i} \neq G_{i+1}$.

In the case where the time is discrete, another possible global representation of
evolution of $\G$ is by the sequence $\SG=G_{1}, G_{2}, \ldots$, where 
$G_i$  corresponds to the static {\em snapshot} of $\G$ at time $t=i$. In this case, it is possible that $G_{i} = G_{i+1}$.

Observe that in both continuous and discrete cases, the underlying graph $G$ (defined in Section~\ref{sec:underlying-graph})  corresponds to the union of all $G_i$ in $\SG$.

The idea of representing a dynamic graph as a sequence of static graphs, mentioned in the conclusion of~\cite{HG97}, was brought to life in~\cite{Fer04} as a combinatorial model called {\em evolving graphs}. An evolving graph usually refers to either one of the two structures $(G,\SG, \ST)$ or $(G,\SG, \mathbb{N})$, the latter used only when discrete-time is considered. Their initial version also included a latency function, which makes them a valid -- graph-centric -- representation of TVGs.

\subsection{Subgraphs of a time-varying graph}

Subgraphs of a TVG $\G$ can be defined in a classical manner, by restricting the set of vertices or edges of $\G$. More interesting is the possibility to define a \emph{temporal subgraph} by restricting the lifetime $\T$ of $\G$, leading to the graph $\G'=(V,E',\T',\rho',\zeta')$ such that 
\begin{itemize}
\item $\T' \subseteq \T$
\item $E' = \{e \in E : \exists t \in \T' : \rho(e,t)=1 \wedge t+\zeta(e,t)\in \T'\}$
\item  $\rho': E' \times \T' \rightarrow \{0,1\}$ where
$\rho'(e,t)=\rho(e,t)$ 
\item 
$\zeta': E' \times \T' \rightarrow \mathbb{T}$ where
$\zeta'(e,t)=\zeta(e,t)$ 
\end{itemize}
In practice, we allow the notation $\G'=\G_{[t_a, t_b)}$ to denote the temporal subgraph of $\G$ restricted to $\T'=\T\cap [t_a, t_b)$, which includes the possible notations $\G_{[t_a, +\infty)}$ or $\G_{(-\infty,t_b)}$ to denote the temporal subgraphs of $\G$ going from $t_a$ to the end of its lifetime, or from the beginning of its lifetime to $t_b$, regardless of whether $\T$ is open, semi-closed, or closed.

\subsection{Journeys}

A sequence of couples $\J=\{(e_1,t_1),$ $(e_2,t_2) \dots,$ $(e_k,t_k)\}$, such that $\{e_1, e_2,...,e_k\}$ is a  walk in $G$ is a {\em journey} in $\G$ if and only if $\rho(e_i,t_i)=1$ and $t_{i+1}\ge t_i + \zeta(e_i,t_i)$ for all $i<k$. Additional constraints may be required in specific domains of application, such as the condition $\rho_{[t_i,t_i+\zeta(e_i,t_i))}(e_i)=1$ in communication networks (the edge remains present until the message is delivered).

 We denote by $departure(\J)$, and $arrival(\J)$, the starting date $t_1$ and the last date $t_k+\zeta(e_k,t_k)$ of a journey $\J$, respectively.
 Journeys can be thought of as {\em paths over time} from a source to a destination and therefore have both a {\em topological} length and a {\em temporal} length.
 The {\em topological length} of $\J$ is the number $\topo{|\J|}= k$ of couples in $\J$ (i.e., the number of {\em hops}); its {\em temporal length} is its end-to-end duration: $arrival(\J) - departure(\J)$. 

Let us denote by $\J^*_\G$ the set of all possible journeys in a time-varying graph $\G$, and by  $\J^*_{(u,v)} \subseteq \J^*_\G$ those journeys starting at node $u$ and ending at node $v$. If a journey exists from a node $u$ to a node $v$, that is, if $\J^*_{(u,v)} \ne \emptyset$, then we say that $u$ can {\em reach} $v$, and allow the simplified notation $u\leadsto v$. Clearly, the existence of journey is not symmetrical: $u \leadsto v \nLeftrightarrow v \leadsto u$; this holds regardless of whether the edges are directed or not, because the time dimension creates its own level of direction. Given a node $u$, the set $\{v\in V : u \leadsto v\}$ is called the {\em horizon} of $u$.

\subsection{Distance}
\label{sec:distance}

As observed, the length of a journey can be measured both in terms of hops or time. This gives rise to two distinct definitions of {\em distance} in a time-varying graph~$\G$:
 \begin{itemize}
  \item The {\em topological distance} from a node $u$ to a node $v$ at time $t$, noted 
  $\topo{d}_{u,t}(v)$, is defined as $Min\{\topo{|\J|}:\J \in \J^*_{(u,v)}, departure(\J) \ge t\}$. For a given date $t$, a journey whose departure is $t'\ge t$ and topological length is equal to $\topo{d}_{u,t}(v)$ is qualified as {\em shortest} ;
  
  \item The {\em temporal distance} from $u$ to $v$ at time $t$, noted $\temp{d}_{u,t}(v)$ is defined as $Min\{arrival(\J):\J \in \J^*_{(u,v)}, departure(\J)\ge t\} - t$. Given a date $t$, a journey whose departure is $t'\ge t$ and arrival is $t+\temp{d}_{u,t}(v)$ is qualified as {\em foremost}. Finally, for any given date $t$, a journey whose departure is $\ge t$ and temporal length is $Min\{\temp{d}_{u,t'}(v) : t' \in \T \cap [t,+\infty)\}$ is qualified as {\em fastest}.
\end{itemize}

The problem of computing shortest, fastest, and foremost journeys in delay-tolerant networks was introduced in~\cite{BFJ03}, and an algorithm for each of the three metrics was provided for the {\em centralized}  version of the problem (assuming complete knowledge of $\G$). Temporal distance and related concepts have been practically used in various fields ranging from social network analysis~\cite{TMML10} to warning delivery protocols in vehicular networks~\cite{RRS11}.

A concept closely related to that of temporal distance is that of {\em temporal view}, introduced in~\cite{KosKW08} in the context of social network analysis. The temporal view\footnote{This concept  was called simply ``view'' in \cite{KosKW08}; since the term {\em view} has a very different meaning in distributed computing (e.g., \cite{YamK96}),  the adjective ``temporal'' has been added to avoid confusion.}
 that a node $v$ has of another node $u$ at time $t$, denoted $\phi_{v,t}(u)$, is defined as the latest (i.e., largest) $t'\le t$ at which a message received by time $t$ at $v$ could have been emitted at $u$; that is, in our formalism,

\begin{center}
  \small
  $\phi_{v,t}(u)=$
  Max$\{departure(\J) : \J \in\J^*_{(u,v)}, arrival(\J)\le t\}$.
\end{center}

The question of knowing whether all the nodes of a network could know their temporal views in real time was recently answered (affirmatively) in~\cite{CasFMS11}. 

\subsection{Other temporal concepts}
\label{sec:other-distance}

The number of definitions built on top of temporal concepts could grow endlessly, and our aim is certainly not to enumerate all of them. Yet, here is a short list of additional concepts that we believe are general enough to be possibly useful in several analytical contexts.

The concept of eccentricity can be separated into a {\em topological eccentricity} and a {\em temporal eccentricity}, following the same mechanism as for the concept of distance. The temporal eccentricity of a node $u$ at time $t$, $\temp{\ecc}_t(u)$, is defined as $max\{\temp{d}_{u,t}(v):v \in V\}$, that is, the duration of the ``longest'' foremost journey from $u$ to any other node. The concept of diameter can similarly be separated into those of {\em topological diameter} and {\em temporal diameter}, the latter being defined at time $t$ as $max\{\temp{\ecc}_t(u):u\in V\}$. These temporal versions of eccentricity and diameter were proposed in~\cite{BFJ03} for the case that the reference time $t$ is the initial time $t_0$ of the system. The temporal diameter was further investigated from a stochastic point of view by Chaintreau {\em et al.} in~\cite{ChMMD08}.

Clementi {\em et al.} introduced in~\cite{CleP10} a concept of {\em dynamic expansion} --~the dynamic counterpart of the concept of {\em node expansion} in static graphs~-- which accounts for the maximal speed of information propagation. Given a subset of nodes $V'\subseteq V$, and two dates $t_1, t_2 \in \T$, the dynamic expansion of $V'$ from time $t_1$ to time $t_2$ is the size of the set $\{v \in V\smallsetminus V' : \exists \J_{(u,v)} \in \J^*_{\G{[t_1,t_2)}} : u \in V'\}$, that is, in a sense, the {\em collective horizon} of $V'$ in $\G_{[t_1,t_2)}$.

The concept of journey was dissociated in~\cite{CasFMS11} into {\em direct} and {\em indirect} journeys. A journey $\J=\{(e_1,t_1),$ $(e_2,t_2) \dots,$ $(e_k,t_k)\}$ is said {\em direct} iff $\forall i, 1\le i < k$, $\rho(e_{i+1},t_i+\zeta(e_i,t_i))=1$, that is, every {\em next} edge in $\J$ is directly available; it is said indirect otherwise. The knowledge of whether a journey is direct or indirect was directly exploited by the distributed algorithm in~\cite{CasFMS11}  to compute temporal distances between nodes. Such a parameter could also play a role in the context of delay-tolerant routing, indicating whether a store-carry-forward mechanism is required (for indirect journeys).

\section{TVG Classes}
\label{sec:classes}

In this section we discuss the impact of properties of TVGs on the feasibility and complexity of distributed problems,  reviewing and unifying existing works from the literature. 
In particular, we identify a hierarchy of  {\em classes} of TVGs based on temporal properties that are formulated using the concepts presented in the previous section. These class-defining properties, organized in an ascending order of assumptions --~from more general to more specific,  are important in that they imply necessary conditions and impossibility
results for distributed computations. 

Let us start with the simplest Class.

\begin{class}
  \label{cl:emitter-possible}
  $\exists u\in V : \forall v\in V, u \leadsto v$.
\end{class}
\noindent That is, at least one node can reach all the others. This condition is   necessary, for example,  
for broadcast to be feasible from at least one node.

\begin{class}
  \label{cl:receiver-possible}
  $\exists u\in V : \forall v\in V, v \leadsto u$.
\end{class}
\noindent That is, at least one node can be reached by all the others. This condition is necessary to be able to compute a function whose input is spread over all the nodes, with at least one node capable of generating the output. Any algorithm for which a terminal state must be causally related to all the nodes initial states also falls in this category, such as leader election in anonymous networks or counting the number of nodes.

\begin{class}\ct{Connectivity over time}
  \label{cl:connectivity}
  $\forall u,v\in V, u \leadsto v$.
\end{class}

\noindent That is, every node can reach all the others; in other words, the TVG is connected over time. By the same discussions as for Class~\ref{cl:emitter-possible} and Class~\ref{cl:receiver-possible}, this condition is necessary to enable broadcast from any node, to compute a function whose output is known by all the nodes, or to ensure that every node has a chance to be elected. 

These three basic classes were used e.g. in~\cite{CasCF09} to investigate how relations between TVGs properties and feasibility of algorithms could be {\em formally} established, based on a combination of evolving graphs~\cite{Fer04} and graph relabelings~\cite{LMS99}. Variants of these classes can be found in recent literature, {\it e.g.} in~\cite{GAS11} where the assumption that connectivity over time eventually takes place among a stable subset of the nodes is used to implement failure detectors in dynamic networks.

\begin{class}\ct{Round connectivity}
  \label{cl:round-connectivity}
  $\forall u,v\in V, \exists \J_1\in \J^*_{(u,v)}, \exists \J_2\in \J^*_{(v,u)} : arrival(\J_1) \le departure(\J_2)$.
\end{class}
\noindent That is, every node can reach all the others and be reached back {\em afterwards}. Such a condition may be required e.g. for adding explicit termination to broadcast, election, or counting algorithms.

The classes defined so far are in general relevant in the case that the lifetime is {\em finite} and a limited number of topological events are considered. When the lifetime is {\em infinite}, connectivity over time is generally assumed on a regular basis, and more elaborated assumptions can be considered.

\begin{class}\ct{Recurrent connectivity}
  \label{cl:recurrent-connectivity}
  $\forall u,v\in V, \forall t\in \T, \exists \J \in \J^*_{(u,v)} : departure(\J) > t$.
\end{class}
\noindent That is, at any point $t$ in time, the temporal subgraph $\G_{[t,+\infty)}$ remains connected over time. This class is implicitly considered in most works on delay-tolerant networks. It indeed represents those DTNs where routing can always be achieved over time. This class was referred to as {\em eventually connected networks} by Awerbuch and Even in \cite{Awerbuch84}, although the terminological compound ``eventually connected'' was also used with different meaning in the recent literature (which we mention in another definition below).

As discussed in Section~\ref{sec:underlying-graph}, the fact that the underlying graph $G=(V,E)$ is connected does not imply that $\G$ is connected over time -- the ordering of topological events matters. Such a condition is however {\em necessary} to allow connectivity over time and thus to perform any type of global computation. Therefore, the following three classes explicitly assume that the underlying graph $G$ is connected.

\begin{class}\ct{Recurrence of edges}
  \label{cl:recurrent-edges}
  $\forall e\in E, \forall t\in \T, \exists t'>t : \rho(e,t')=1$  and $G$ is connected.
\end{class}

\noindent That is, if an edge appears once, it appears infinitely often. Since the underlying graph $G$ is connected, we have Class {\ref{cl:recurrent-edges}} $\subseteq$ Class {\ref{cl:recurrent-connectivity}}. Indeed, if all the edges of a connected graph appear infinitely often, then there must exist, by transitivity, a journey between any pairs of nodes infinitely often. 

In a context where connectivity is recurrently achieved, it becomes interesting to look at problems where more specific properties of the journeys are involved, e.g. the possibility to broadcast a piece of information in a shortest, foremost, or fastest manner (see Section~\ref{sec:distance} for definitions). Interestingly, these three declinations of the same problem have different requirements in terms of TVG properties. It is for example possible to broadcast in a foremost fashion in Class~$\ref{cl:recurrent-edges}$, whereas shortest and fastest broadcasts are not possible~\cite{CasFMS10}. 

Shortest broadcast becomes however possible if the recurrence of edges is bounded in time, and the bound known to the nodes, a property characterizing the next class:

\begin{class}\ct{Time-bounded recurrence of edges}
  \label{cl:bounded-recurrent-edges}
  $\forall e \in E, \forall t \in \T, \exists t' \in [t, t+\Delta), \rho(e,t')=1$, for some $\Delta \in \mathbb{T}$
    and $G$ is connected.
\end{class}

\noindent Some implications of this class include a temporal diameter that is bounded by $\Delta Diam(G)$, as well as the possibility for the nodes to wait a period of $\Delta$ to discover all their neighbors (if $\Delta$ is known). The feasibility of shortest broadcast follows naturally by using a $\Delta$-rounded breadth-first strategy that minimizes the topological length of journeys.

A particular important type of bounded recurrence is the periodic case:

\begin{class}\ct{Periodicity of edges}
  \label{cl:periodic-edges}
  $\forall e\in E, \forall t\in \T, \forall k \in \mathbb{N}, \rho(e,t)=\rho(e,t+kp)$, for some $p \in \mathbb{T}$   and $G$ is connected.
\end{class}

\noindent The periodicity assumption holds in practice in many cases, including networks whose entities are mobile with periodic movements (satellites, guards tour, subways, or buses). The periodic assumption within a delay-tolerant network has been considered, among others, in the contexts of network exploration~\cite{FMS09,IW11,FKMS12} and routing~\cite{KerO09,LW09b}. Periodicity enables also the construction of foremost broadcast trees that can be re-used (modulo $p$ in time) for subsequent broadcasts~\cite{CasFMS11} (whereas the more general classes of recurrence requires the use of a different tree for every foremost broadcast).

More generally, the point in exploiting TVG properties is to rely on invariants that are generated by the dynamics ({\it e.g.} recurrent existence of journeys, periodic optimality of a broadcast tree, {\it etc.}). In some works, particular assumptions on the network dynamics are made to obtain invariants of a more classical nature. Below are some examples of classes, formulated using the graph-centric point of view of (discrete-time) evolving graphs, {\it i.e.}, where $\G=(G,\SG,\mathbb{N})$.
\begin{class}\ct{Constant connectivity}
  \label{cl:always-connected}
  $\forall G_i \in \SG, G_i$ is connected.
\end{class}
\noindent Here, the dynamics of the network is not constrained as long as it remains connected in every time step. Such a class was used for example in~\cite{OW05} to enable progression hypotheses on the broadcast problem. Indeed, if the network is always connected, then at every time step there must exist an edge between an informed node and a non-informed node, which allows to bound broadcast time by $n=|V|$ time steps (worst case scenario). This class was also considered in~\cite{KMO11} for the problem of consensus.

\begin{class}\ct{T-interval connectivity}
  \label{cl:t-interval-connected}
$\forall i \in \mathbb{N}, T \in \mathbb{N}, \exists G' \subseteq G : V_{G'}=V_G, G'$ is connected, and $\forall j \in [i,i+T-1), G' \subseteq G_j$.
\end{class}
\noindent This class is a particular case of constant connectivity in which a same spanning connected subgraph of the underlying graph $G$ is available for any period of $T$ consecutive time steps. It was introduced in~\cite{KLO10} to study problems such as counting, token dissemination, and computation of functions whose input is spread over all the nodes (considering an adversarial edge schedule). The authors show that computation could be sped up of a factor $T$ compared to the 1-interval connected graphs, that is, graphs of Class~\ref{cl:always-connected}. 

Other classes of TVGs can be found in~\cite{RBK07}, based on intermediate properties between constant connectivity and connectivity over time. They include Class~\ref{cl:eventually-connected} and Class~\ref{cl:eventually-routable} below.
\begin{class}\ct{Eventual instant-connectivity}
  \label{cl:eventually-connected}
  $\forall i\in \mathbb{N}, \exists j \in \mathbb{N} : j \ge i$, $G_j$ is connected. In other words, there is always a future time step in which the network is instantly connected. 

This class was simply referred to as {\em eventual connectivity} in~\cite{RBK07}, but since the meaning is different than that of~\cite{Awerbuch84} (connectivity over time), we renamed it to avoid ambiguities.
\end{class}

\begin{class}\ct{Eventual instant-routability}
  \label{cl:eventually-routable}
  $\forall u,v \in V, \forall i\in \mathbb{N}, \exists j \in \mathbb{N} : j \ge i$ and a path from $u$ to $v$ exists in $G_j$.
\end{class}

\noindent  That is, for any two nodes, there is always a future time step in which a instant path exists between them. The difference with Class~\ref{cl:eventually-connected} is that these paths may occur at different times for different pairs of nodes. 
Classes \ref{cl:eventually-connected} and  \ref{cl:eventually-routable} were used in~\cite{RBK07} to represent networks where routing protocols for (connected) mobile ad hoc networks eventually succeed if they tolerate transient topological faults.

Most of the works listed above strove to characterize the impact of various temporal properties on problems or algorithms. A reverse approach was considered by Angluin {\em et al.} in the field of {\em population protocols}~\cite{AAD+06,AAER07}, where for a given assumption (that any pair of node interacts infinitely often), they characterized all the problems that could be solved in this context. The corresponding class is generally referred to as that of {\em (complete) graph of interaction}.

\begin{class}\ct{Complete graph of interaction}
  \label{cl:complete}
  The underlying graph $G$$=$$(V,E)$ is complete, and $\forall e\in E, \forall t\in \T, \exists t'>t : \rho(e,t')$$=$$1$.
\end{class}

\noindent From a time-varying graph perspective, this class is the specific subset of Class~\ref{cl:recurrent-edges}, in which the underlying graph $G$ is complete. Various types of schedulers and assumptions have been subsequently considered in the field of population protocols, adding further constraints to Class~\ref{cl:complete} (e.g. weak fairness, strong fairness, bounded, or k-bounded schedulers) as well as interaction graphs which might not be complete.

\begin{figure}
  \centering
\begin{tikzpicture}[sloped, level distance=34pt, font=\small, sibling distance=30pt]
  \tikzstyle{every path}=[<-]
  \node {${\cal C}_{\ref{cl:round-connectivity}}$}
  child [grow=right]{
    node {${\cal C}_{\ref{cl:connectivity}}$}
    edge from parent [->]
    child {
      node {${\cal C}_{\ref{cl:receiver-possible}}$}
      edge from parent [->]
    }
    child {
      node {${\cal C}_{\ref{cl:emitter-possible}}$}
      edge from parent [->]
    }
  }
  child [grow=left]{
    node {${\cal C}_{\ref{cl:recurrent-connectivity}}$}
    child {
      node {${\cal C}_{\ref{cl:recurrent-edges}}$}
      child {
        node {${\cal C}_{\ref{cl:complete}}$}
      }
      child [sibling distance=16pt] {
        node {${\cal C}_{\ref{cl:bounded-recurrent-edges}}$}
        child {
          node {${\cal C}_{\ref{cl:periodic-edges}}$}
        }
      }
    }
    child {
      node {${\cal C}_{\ref{cl:eventually-routable}}$}
      child {
        node {${\cal C}_{\ref{cl:eventually-connected}}$}
        child {
          node {${\cal C}_{\ref{cl:always-connected}}$}
          child {
            node {${\cal C}_{\ref{cl:t-interval-connected}}$}
          }
        }
      }
    }
  };
\end{tikzpicture}
\caption{\label{fig:classification} Relations of inclusion between classes ({\it from specific to general}).}
\end{figure}
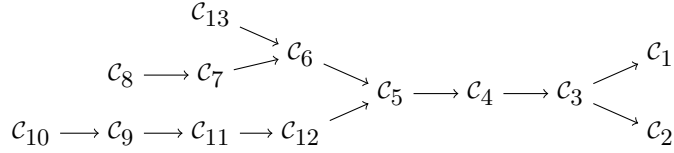

An interesting aspect of unifying these properties within the same formalism is the possibility to see how they relate to one another,  and to compare the associated solutions or algorithms. 
An insight for example can be gained  by  looking  at the short classification 
shown in Figure~\ref{fig:classification}, 
where basic relations of inclusion between the above classes are reported. These inclusion are {\em strict}:   for each relation, the parent class contains some 
time-varying graphs that are not in the child class.

Clearly, one should try to solve a problem in the most general context possible. The right-most classes are so general that they offer little properties to be exploited by an algorithm, but some intermediate classes, such as Class~$\ref{cl:recurrent-connectivity}$,  appear quite central in the hierarchy. This class indeed 
contains all the classes where significant work was done. A problem solved in this class would therefore apply to virtually all the contexts considered heretofor in the literature.

Such a classification may also be used to categorize problems themselves. As mentioned above, shortest broadcast is not generally achievable  in Class~\ref{cl:recurrent-edges}, whereas foremost broadcast is. Similarly, it was shown in~\cite{CasFMS10} that 
fastest broadcast is not feasible in Class~\ref{cl:bounded-recurrent-edges}, whereas shortest broadcast
can be achieved with some knowledge. Since Class~\ref{cl:bounded-recurrent-edges} $\subset$ Class~\ref{cl:recurrent-edges}, we have 
\begin{center}
$foremostBcast\preceq shortestBcast\preceq fastestBcast$ \end{center}
where $\preceq$ is a partial order on these problems topological requirements.

\section{TVG and Network Analysis}
\label{sec:analysis}

This section is concerned with the {\it a posteriori} analysis of network traces. We discuss three particular aspects of this general question, which are i) how network traces could be checked for inclusion in some of the above classes, ii) how temporal concepts can be leveraged to express new phenomenon or properties in complex systems, and iii) how TVGs could be used to study a coarser-grain evolution of network properties, whether these properties are of a classical or a temporal nature (which implies different approaches).

\subsection{Recognizing TVGs}

\begin{figure}[h]
  \centering
  \begin{tabular}{c|c|c}
  \begin{tikzpicture}[xscale=.62]
    \tikzstyle{every node}=[draw, fill, circle, inner sep=1pt]
    \path (0,0) node (a) {};
    \path (a)+(-.6,-.2) coordinate (a');
    \path (3,0) node (b) {};
    \path (b)+(.6,.2) coordinate (b');
    \path (2.5,1) node (c) {};
    \path (c)+(-.5,0) coordinate (c');
    \path (0.5,-1) node (d) {};
    \path (d)+(.5,0) coordinate (d');
    \tikzstyle{every node}=[]
    \path[left] (a) node {$a$};
    \path[right] (b) node {$b$};
    \path[above] (c) node {$c$};
    \path[below] (d) node {$d$};
    \tikzstyle{every path}=[draw, semithick]
    \path (a)--(d);
    \path (b)--(c);
    \tikzstyle{every path}=[draw, ->]
    \path (c)--(c');
    \path (d)--(d');
    \path[rounded corners=3pt] (a') rectangle (b');
  \end{tikzpicture}
  &
  \begin{tikzpicture}[xscale=.62]
    \tikzstyle{every node}=[draw, fill, circle, inner sep=1pt]
    \path (0,0) node (a) {};
    \path (a)+(-.6,-.2) coordinate (a');
    \path (3,0) node (b) {};
    \path (b)+(.6,.2) coordinate (b');
    \path (1.5,1) node (c) {};
    \path (c)+(-.5,0) coordinate (c');
    \path (1.5,-1) node (d) {};
    \path (d)+(.5,0) coordinate (d');
    \tikzstyle{every node}=[]
    \path[left] (a) node {$a$};
    \path[right] (b) node {$b$};
    \path[above] (c) node {$c$};
    \path[below] (d) node {$d$};
    \tikzstyle{every path}=[draw, ->]
    \path (c)--(c');
    \path (d)--(d');
    \path[rounded corners=3pt] (a') rectangle (b');
  \end{tikzpicture}
  &
  \begin{tikzpicture}[xscale=.62]
    \tikzstyle{every node}=[draw, fill, circle, inner sep=1pt]
    \path (0,0) node (a) {};
    \path (a)+(-.6,-.2) coordinate (a');
    \path (3,0) node (b) {};
    \path (b)+(.6,.2) coordinate (b');
    \path (.5,1) node (c) {};
    \path (2.5,-1) node (d) {};
    \tikzstyle{every node}=[]
    \path[left] (a) node {$a$};
    \path[right] (b) node {$b$};
    \path[above] (c) node {$c$};
    \path[below] (d) node {$d$};
    \tikzstyle{every path}=[draw, semithick]
    \path (a)--(c);
    \path (b)--(d);
    \tikzstyle{every path}=[draw]
    \path[rounded corners=3pt] (a') rectangle (b');
  \end{tikzpicture}
  \\
  at date $t_1$&at date $t_2>t_1$&at date $t_3>t_2$
  \end{tabular}
  \caption{\label{fig:open-tcc}Example of \emph{connected component} (\textit{here, $\{a,b\}$ using external edges).}}
\end{figure}
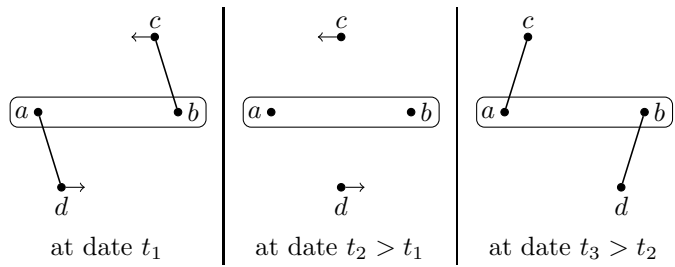

Let us start with recognition of properties that relate to connectivity. In~\cite{BF03}, the problem of computing connected-component in a given evolving graph $\G$ is considered. A {\em (time-)connected component} in $\G$ is therein defined as a set of nodes $V'\subseteq V$ such that $\forall u,v \in V', u \leadsto v$. The authors observe that in general, some of the journeys' edges that contribute to a component may involve nodes {\em outside} the component, as illustrated in Figure~\ref{fig:open-tcc}. Variations around the concept of connected component include for example {\em strongly}-connected components in~\cite{BF03}, as well as {\em in-} and {\em out-}components in~\cite{TMML10}.

\begin{figure}[h]
\def\wgraph (#1,#2){%
  \tikzstyle{every node}=[draw,fill,circle,inner sep=#1]
  \path (0,0) node (c){};
  \path (c)+(-1,.4) node (a){};
  \path (c)+(1,.4) node (e){};
  \path (c)+(-.6,-.6) node (b){};
  \path (c)+(.6,-.6) node (d){};
  \tikzstyle{every node}=[font=\footnotesize]
  \path (a)+(-#2,0) node (la){$a$};
  \path (b)+(-#2,0) node (lb){$b$};
  \path[above] (c) node (lc){$c$};
  \path (d)+(#2,0) node (ld){$d$};
  \path (e)+(#2,0) node (le){$e$};
}
\centering
  \begin{tikzpicture}[scale=2]
    \wgraph(.8pt,4pt)
    \tikzstyle{every node}=[sloped,below,font=\scriptsize,inner sep=2pt]
    \draw (a)--node{$[1,2)$}(b);
    \draw (b)--node{$[2,3)$}(d);
    \draw (d)--node{$[3,5)$}(e);
    \tikzstyle{every node}=[sloped,above,font=\scriptsize,inner sep=1pt]
    \draw (a)--node[inner sep=1.5pt]{$[1,2)$}(c);
    \draw (b)--node[inner sep=1.5pt]{$[2,3)$}(c);
    \draw (c)--node{$[2,4)$}(d);
    \draw (c)--node{$[3,5)$}(e);
    
    \tikzstyle{every node}=[draw, fill, inner sep=.1pt]
    \path (2.6,-.1) coordinate (zz);
    \path (zz)+(18:.6) node (ee) {};
    \path (zz)+(90:.6) node (cc) {};
    \path (zz)+(162:.6) node (aa) {};
    \path (zz)+(234:.6) node (bb) {};
    \path (zz)+(306:.6) node (dd) {};
    \tikzstyle{every node}=[]
    \path (aa)+(-.1,.02) node (laa){$a$};
    \path (bb)+(-.08,-.08) node (laa){$b$};
    \path (cc)+(0,.09) node (laa){$c$};
    \path (dd)+(.08,-.08) node (laa){$d$};
    \path (ee)+(.1,.02) node (laa){$e$};
    
    \tikzstyle{every path}=[>=stealth, shorten >=1.5pt, shorten <=1.5pt]
    \path[<->] (aa) edge [bend right=30] (bb);
    \path[<->] (aa) edge [bend left=30] (cc);
    \path[->] (aa) edge [bend right=10] (dd);
    \path[->] (aa) edge [bend left=10] (ee);
    \path[<->] (bb) edge [bend left=10] (cc);
    \path[<->] (bb) edge [bend right=30] (dd);
    \path[->] (bb) edge [bend right=10] (ee);
    \path[<->] (cc) edge [bend left=10] (dd);
    \path[<->] (cc) edge [bend left=30] (ee);
    \path[<->] (dd) edge [bend right=30] (ee);
    
    \path (zz)+(-1.4,0) coordinate (zzleft);
    \draw[->, semithick, shorten >=50pt] (zzleft)--(zz);
  \end{tikzpicture}
\caption{\label{fig:closure}Transitive closure of journeys (example from~\cite{CasCF09})}
\end{figure}
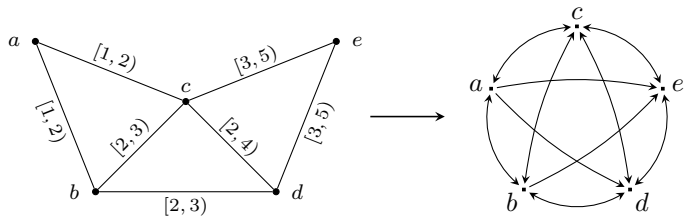

The problem of determining the largest connected component in a given TVG $\G$ was shown NP-complete in~\cite{BF03}, through reduction from the {\em maximum clique} problem. Yet, checking whether a given set of nodes is a connected component in $\G$ can be done easily provided a few transformations, as described in the same paper. Consider the \emph{transitive closure} of all journeys of a graph $\G$, given as the graph $H = (V,A_{H})$, where $A_{H} = \{(u,v) : u \leadsto v\}$. The transitive closure is a static and directed graph, as illustrated in Figure~\ref{fig:closure}, since journeys are by nature directed entities. The computation of transitive closures can be done efficiently by building a tree of \emph{shortest} journeys for each node, using any of the algorithms in~\cite{BFJ03}.

Checking whether a set of nodes is a connected component in $\G$ now comes to check whether it is a clique in $H$ (which is easy). The concept of transitive closure actually allows to check a graph $\G$ for inclusion in several classes. For instance, the graph is in Class~\ref{cl:emitter-possible} iff $H$ possesses an out-dominating set of size $1$; it is in Class~\ref{cl:receiver-possible} iff $H$ possesses an in-dominating set of size $1$; it is in Class~\ref{cl:connectivity} iff $H$ is a complete graph. The reader is referred to~\cite{CasCF09} for more examples of inclusion checking, based on other classes and transformations.

\subsection{Transposing the definition of phenomena}
The following paragraphs discuss the possible use of TVGs to express the redefinition (or {\em translation}) of usual concepts in complex system analysis, into a dynamic version. We provide two examples: the {\em small world} effect, and the {\em fairness} in a network. Further examples could certainly be found.

\subsubsection{Small World}
A small-world network is one where the distance between two randomly chosen nodes (in terms of hops) grows logarithmically with the number of nodes in the network.
Time-varying graph concepts, such as those of journeys, connectivity over time, and temporal distance have been used in~\cite{TSM+09} to characterize the small world behavior of real-world networks in temporal terms, that is, the fact that there is always a journey of short {\em duration} between any two nodes. Among the concepts introduced in~\cite{TSM+09} is the {\em characteristic temporal path length}, defined as

\begin{center}
  \large
  $\frac{\sum_{u,v \in V} \temp{d}_{t_0}(u,v)}{|V^2|}$
\end{center}

\noindent where $t_0$ is the first date in the network lifetime $\T$. In other words, this value is the average of temporal distances between all pairs of nodes at starting time. An average of this value over the network lifetime would certainly be meaningful as well. 

As per the {\em topological} meaning ({\it i.e.,} in terms of hops) of the small world property in a dynamic context, e.g. the fact that ``mobile networks have a diameter of 7''~\cite{PS00}, it could be formalized as follows:
\begin{center}
$\forall u,v \in V, \forall t\in \T, \exists \J \in \J^*_{(u,v)} : departure(\J)\ge t , \topo{|\J|} \le 7$.
\end{center}

\subsubsection{Fairness and Balance}
\label{sec:fairness}

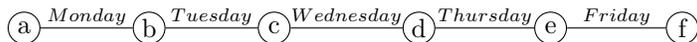
\begin{figure}[h]
  \begin{center}
    \begin{tikzpicture}[xscale=1.75]
      \tikzstyle{every node}=[draw,circle, minimum size=12pt, inner sep=0pt]
      \path (0,0) node (a){a};
      \path (a)+(.95,0) node (b){b};
      \path (a)+(1.9,0) node (c){c};
      \path (a)+(3,0) node (d){d};
      \path (a)+(4,0) node (e){e};
      \path (a)+(5,0) node (f){f};
      \tikzstyle{every node}=[below,font=\scriptsize,inner sep=1pt]
      \draw (a)--node[above]{$Monday$}(b);
      \draw (b)--node[above]{$Tuesday$}(c);
      \draw (c)--node[above]{$Wednesday$}(d);
      \draw (d)--node[above]{$Thursday$}(e);
      \draw (e)--node[above]{$Friday$}(f);
       \end{tikzpicture}
    \caption{\label{fig:ex1} Weekly interactions between six people.}
  \end{center}
\end{figure}

Other properties of interest can take the form of quantities or statistical information. 
Consider the caricatural example of Figure~\ref{fig:ex1}, where nodes $a$ to $f$ represent individuals, each of which meets some other individuals every week (on a periodical basis). 

A glance at the structure of this network does not reveal any strong anomaly: a) the graph is a line, b) its diameter is $5$, c) nodes $c$ and $d$ are more central than the other nodes, {\it etc}. 
However, if we consider the temporal dimension of this graph, it appears that (the interaction described by) the graph is highly unfair and asymmetric: any  information originating from $a$ can reach $f$ within $5$ to $11$ days (depending on what day it is originated), whereas information from $f$ needs about one month to reach $a$. 
Node $a$ also appears more central than $c$ and $d$ from a temporal point of view.

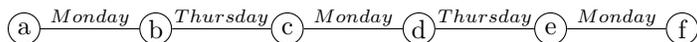
\begin{figure}[h]
  \begin{center}
    \begin{tikzpicture}[xscale=1.75]
      \tikzstyle{every node}=[draw,circle, minimum size=12pt, inner sep=0pt]
      \path (0,0) node (a){a};
      \path (a)+(1,0) node (b){b};
      \path (a)+(2,0) node (c){c};
      \path (a)+(3,0) node (d){d};
      \path (a)+(4,0) node (e){e};
      \path (a)+(5,0) node (f){f};
      \tikzstyle{every node}=[below,font=\scriptsize,inner sep=1pt]
      \draw (a)--node[above]{$Monday$}(b);
      \draw (b)--node[above]{$Thursday$}(c);
      \draw (c)--node[above]{$Monday$}(d);
      \draw (d)--node[above]{$Thursday$}(e);
      \draw (e)--node[above]{$Monday$}(f);
    \end{tikzpicture}
    \caption{\label{fig:ex3} Weekly interactions between six people - Fairer version.}
  \end{center}
\end{figure}

We could define here a concept of {\em fairness} as being the standard deviation among the nodes temporal eccentricities (see Section~\ref{sec:other-distance}). This indicator provides an outline on how well the interactions are balanced among nodes. 
For instance, the TVG of Figure~\ref{fig:ex1} is highly unfair; while the one shown in Figure~\ref{fig:ex3} is fairer (still, the fairness remains strucurally constrained by $G$, the underlying graph).

A related measure could reflect how balanced the graph is with respect to the time dimension, since the metrics of interest are time-dependent ({\it e.g.} the temporal diameter of the TVG in Figure~\ref{fig:ex3} is much lower on Mondays than on Tuesdays). Recent efforts in similar directions include measuring the temporal distance between individuals based on e-mail datasets~\cite{KosKW08, Kostakos09} or inter-meeting times~\cite{TSM+09}, or the redefinition of further concepts built on top of temporal distance, such as {\em temporal betweenness} and {\em temporal closeness}~\cite{TMMLN10}.

\subsection{Capturing the coarse-grain evolution}

At many occasions in this paper, we have focused on the question of how static concepts translate into a dynamic context, e.g. through the redefinition of more basic notions like those of paths (into journeys), distance (into temporal distance) or connectivity (into connectivity over time). From a complex system perspective, these {\em temporal} indicators, as well as those built on top of them, are completing the set of {\em atemporal} indicators usually considered, such as (the normal versions of) distance and diameter, density, clustering coefficient, or modularity, to name a few. It is important to keep in mind that all these indicators, whether temporal or atemporal, essentially accounts for network properties at a reasonably short time-scale ({\em fine-grain} dynamics). They do not reflect how network evolves over longer periods of time ({\em coarse-grain} dynamics). 

We present below a general approach to look at the evolution of both atemporal and temporal indicators  \cite{SQFCA11}. Looking at the evolution of atemporal indicators can be done by representing the evolution of the network as a sequence of {\em static} graphs, each of which represents the aggregated interactions over a given time-window. The usual indicators can then be normally measured on these graphs and their evolution studied over time. The case of temporal indicators is more complex because the corresponding evaluation cannot be done on static graphs. The proposed solution is therefore to look at the evolution of temporal indicators through a {\em sequence} of shorter (and non-aggregated) {\em time-varying graphs}, {\it i.e.,} a sequence of {\em temporal subgraphs} of the original time-varying graph that cover successive time-windows. 

\subsubsection{Evolution of Atemporal  Indicators} \label{static}
\label{sec:atemporal}

\paragraph{TVGs as a sequence of footprints}

Given a TVG \TVG, one can define the {\em footprint} of this graph from $t_1$ to $t_2$ as the static graph $G^{[t_1,t_2)}=(V,E^{[t_1,t_2)})$ such that $\forall e \in E, e \in E^{[t_1,t_2)} \iff \exists t \in [t_1,t_2), \rho(e,t)=1$. In other words, the footprint aggregates all interactions of given time windows into static graphs. Let the lifetime  $\T$ of the time-varying graph be partitioned in consecutive sub-intervals $\tau = [t_0,t_1), [t_1,t_2) \ldots [t_i,t_{i+1}), \ldots$; where each $[t_k,t_{k+1})$ can be noted $\tau_k$. We call {\em sequence of footprints} of $\G$ according to $\tau$ the sequence {\em SF}$(\tau) = G^{\tau_0},G^{\tau_1}, \ldots$.
Considering this sequence with a sufficient size of the intervals allows to overcome the strong fluctuations of fine-grain interactions, and focus instead on more general trends of evolution. Note that the same approach could be considered with a sequence of intervals that are overlapping ({\it i.e.,} a {\em sliding} time-window) instead of disjoint ones. Another variation may be considered based on whether the set of nodes in each $G^{\tau_i}$ is also allowed to vary.
Since every graph in the sequence is static,  any classical network parameter can be directly measured on it. 
Depending on the parameter and on the application, different choices of 
granularity are more appropriate to capture a meaningful behavior. At one extreme, each interval could correspond to the smallest time unit (in discrete-time systems), or to the time between any two consecutive modification of the graph; in these cases the whole sequence becomes equivalent to the model of {\em evolving graph}~\cite{Fer04}. 
At the other side of the spectrum, i.e. taking $\tau = \T$, the sequence would consist of a single footprint aggregating all interactions over the network lifetime, that is, be equal to $G$, the underlying graph of $\G$. 

Looking at the evolution of atemporal parameters allows to understand how some emerging phenomena occur on the network structure, for instance, the densification of transportation networks (through diameter or average distance indicators), or the formation of communities in social networks (through {\em modularity}~\cite{BGLL08}, {\em cohesion}~\cite{FCF11} or other indicators e.g.~\cite{ADBV08,BSV07}).

\subsubsection{Evolution of Temporal Indicators} 
\label{sec:temporal}

Most temporal concepts --~including those mentioned in Section~\ref{sec:concepts}~-- are based on replacing the notion of {\em path} by that of {\em journey}. As a result, they can be declined into three versions depending on the type of metric considered ({\it i.e.,} shortest, foremost, fastest).  Since journeys are paths over  {\em time}, the  evolution of parameters based on journeys {\em cannot} be studied using a sequence of aggregated static graphs. For example,  there might be a path  between $x$ and $y$ in all footprints, and yet possibly no journey between them depending on the precise chronology of interaction. Analyzing the evolution of such parameters requires more than a sequence of static graphs. 

\paragraph{TVGs as a sequence of (shorter) TVGs}
Temporal subgraphs have been defined in Section~\ref{sec:concepts}. Roughly speaking, they are themselves TVGs that reproduce all the interactions present in the original TVG for a given time window -- without aggregating them. In the same way as for the sequence of footprints, we can now look at the evolution of a TVG through a sequence of shorter TVGs {\em ST}$(\tau) = {\cal G}^{\tau_0},{\cal G}^{\tau_1}, \ldots$, in which the intervals are either disjoint or overlapping. 
Looking at the coarse-grain evolution of temporal indicators could allow to answer questions like: how does the temporal distance between nodes evolve over time? Or more generally how a network self-organizes, optimizes, or deteriorate, in terms of temporal efficiency. Using concepts like the fairness, defined above, this may also help capture the emergence of non-apparent inequalities in a social network.

\section{Random TVGs}
\label{sec:stochastic}

Randomness in time-varying graphs can be introduced at several different levels.
The most direct one is clearly that provided by  {\em probabilistic time-varying graphs}, where
the 
 presence function  $\rho: E \times \T \rightarrow [0,1]$  indicates  the {\em probability}
that a given edge is available at a given time. 
In a context of mobility, the probability distribution of $\rho$ is intrinsically related to the expected mobility of the nodes.
 Popular example of random mobility models include the 
{\em Random Waypoint} and {\em Random Direction} models \cite{CBD02},
where waypoints of consecutive movements are chosen 
uniformly at random. 
Mobility models from social networks include the
 Time-Variant Community model 
(TVC) \cite{HSPH07}, and  the more recent
Home-cell Community-based Mobility Model
(HCMM) \cite{BoP10}.

Definitions of {\em random} TVG differ  depending on whether  the time is discrete or
continuous.

A {\em (discrete-time) random time-varying graph} is a TVG whose lifetime is an interval  of $\mathbb{N}$ and whose {\em sequence of characteristic graphs} $\SG=G_{1}, G_{2},..$ is such that every $G_i$ is a Erd\"os and R\'enyi random graph; that is, $\forall e\in V^2,\mathbb{P}[e\in E_{G_i}]=p$ for some $p$; this definition is introduced by Chaintreau {\it et al.}~\cite{ChMMD08}.

One particularity of  discrete-time random TVGs is that the $G_i$s are independent with respect to each other. While this definition allows purely random graphs, it does not capture some properties of real world networks, such as the fact that an edge may be more likely to be present in $G_{i+1}$ if it is already present in $G_i$.
This question is addressed  by Clementi {\it et al.}~\cite{CleMMPS08}
  by introducing {\em Edge-Markovian Evolving Graphs}. These are discrete-time evolving graphs in which the presence of every edge follows an individual Markovian process. More precisely, the sequence of characteristic graph $\SG=G_1, G_2,..$ is such that 
\begin{center}
  $
  \begin{cases}
    \mathbb{P}[e\in E_{G_{i+1}}|e\notin E_{G_i}]=p\\
    \mathbb{P}[e\notin E_{G_{i+1}}|e\in E_{G_i}]=q
  \end{cases}
  $
\end{center}
for some $p$ and $q$ called {\em birth rate} and {\em death rate}, respectively. The probability that a given edge remains absent or present from $G_i$ to $G_{i+1}$ is obtained by complement of $p$ and $q$. The very idea of considering a {\em Markovian} Evolving Graph seems to have appeared in~\cite{AKL08}, in which the authors consider a particular case that is substantially equivalent to the discrete-time random TVG from~\cite{ChMMD08}. Variations around the model of edge-markovian evolving graphs include cases where $G_{i+1}$ depends not only on $G_{i}$, but also on older graphs $G_{i-1}$, $G_{i-2}$,... (the edges follow a higher order Markovian process)~\cite{GP10}. Edge-Markovian EGs were used in~\cite{CleMMPS08}, along with the concept of dynamic expansion (see Section~\ref{sec:other-distance}) to address stochastic questions such as {\it does dynamics necessarily slow down a broadcast?} Or {\it can random node mobility be exploited to speed-up information spreading?} Baumann {\it et al.} extended this work in~\cite{BCF09} by establishing tight bounds on the propagation time for any birth and death rates.

A {\em continuous-time random time-varying graph} is  a TVG in which the appearance of every edge obeys a Poisson process, that is, $\forall e\in V^2, \forall t_i \in App(e), \mathbb{P}[t_{i+1}-t_i<d]=\lambda e^{\lambda x}$ for some $\lambda$;  this definition is introduced by Chaintreau {\it et al.}  in~\cite{ChMMD08}.\footnote{It is interesting to note that in their definition of random TVG, the authors rely on a graph-centric point of view in discrete time and on an edge-centric point of view in continuous time.  The same trend can actually be observed in most of the works we reviewed here.}

 Random time-varying graphs, both discrete-time and continuous-time, 
were used in~\cite{ChMMD08} to characterize phase transitions between no-connectivity and connectivity over time as a function of the number of nodes, a given time-window duration, and constraints on both the topological and temporal lengths of journeys.


\section{Research Problems and Directions}

The first most
 obvious research task  is that of exploring the universe of
dynamic networks using the formal tools provided by the TVG formalism.
The long-term goal is that of providing a comprehensive map of this universe,
identifying both the commonality and the natural differences between the
various types of dynamical systems modeled by TVG.
Additionally,  several, more specific research areas can be identified
including the ones described below.

{\em Distributed TVG algorithms design and analysis.}
The design and analysis of distributed algorithms and protocols
for time-varying graphs is an open research area. In fact very few
problems have been attacked so far: routing and broadcasting 
in delay-tolerant networks;  broadcasting and exploration in 
opportunistic-mobility networks; new self-stabilization techniques (such as the one in~\cite{KY05}); detection of emergence and resilience of communities, and viral marketing in social networks.

{\em Design and optimization of TVG.}
If the interactions in a network can be planned -- decided by a designer --, then a number of new interesting optimization problems arise with the design of time-varying graph. They may concern for example the minimization of the temporal diameter or the balancing of nodes eccentricities. E.g. how to modify the days of meeting in Figure~\ref{fig:ex1} so as to minimize the network diameter (still preventing two meetings the same day for each people)? Figure~\ref{fig:ex3} showed a basic improvement (the diameter was between $24$ and $30$ days, and shortened to between $14$ and $20$ days). Is a given setting optimal? How to prove it? What if the underlying graph can also be modified? {\it etc.} A whole field is opening that promizes exciting research avenues.





{\em Complexity Analysis.}
Analyzing the complexity of a distributed algorithm in a TVG -- e.g. in number of messages -- is not trivial, partly because contrarily to the static cases, the complexity of an algorithm in a dynamic network has a strong dependency, not only on the usual network parameters (number of nodes, edges, etc.), but also on the number of topological events taking place during its execution. In many of the algorithms we have encountered, the majority of messages is in fact directly triggered by topological events, e.g., in reaction to the local appearance or disappearance of an edge. The number of topological events therefore represents a new complexity parameter, whose impact on various problems remains to study.

{\em Patterns Detection and Visualization.}
In order to better understand complex systems and their dynamic aspects, data need to be visualized in a way that allows the intuition to guess a
particular property or interaction pattern. Several works are
progressing in this direction, including the {\em Gephi} project~
\cite{Gephi} or the {\em Graphstream} library~\cite{Graphstream}, where
both nodes and edges can be specified with temporal and spatial
attributes that enable the visualization of their evolution. 
These can be used to have a global vision of the phenomena to be
explained at the micro, meso and macro levels. In addition, through
the use of the interaction-centric point view, TVGs enable to look at the interplay between topological aspects that allow local interaction to have global
effects.


\begin{thebibliography}{10}
\small

\bibitem{ADBV08}
J.~I. Alvarez-Hamelin, L.~Dall'Asta, A.~Barrat, and A.~Vespignani.
\newblock K-core decomposition of internet graphs: hierarchies, self-similarity
  and measurement biases.
\newblock {\em Networks and Heterogeneous Media}, 3(2):371--293, 2008.

\bibitem{AAD+06}
D.~Angluin, J.~Aspnes, Z.~Diamadi, M.~Fischer, and R.~Peralta.
\newblock Computation in networks of passively mobile finite-state sensors.
\newblock {\em Distributed Computing}, 18(4):235--253, 2006.

\bibitem{AAER07}
D.~Angluin, J.~Aspnes, D.~Eisenstat, and E.~Ruppert.
\newblock The computational power of population protocols.
\newblock {\em Distributed Computing}, 20(4):279--304, 2007.

\bibitem{AKL08}
C.~Avin, M.~Koucky, and Z.~Lotker.
\newblock {How to explore a fast-changing world}.
\newblock In {\em Proceedings of 35th International Colloquium on Automata, Languages
  and Programming (ICALP)}, pages 121--132, 2008.

\bibitem{Awerbuch84}
B.~Awerbuch and S.~Even.
\newblock Efficient and reliable broadcast is achievable in an eventually connected network.
\newblock In {\em Proceedings of 3rd ACM symposium on Principles of Distributed Computing (PODC'84)}, pages 278--281, 1984.

\bibitem{BHKL06}
L.~Backstrom, D.~Huttenlocher, J.~Kleinberg, and X.~Lan.
\newblock Group formation in large social networks: membership, growth, and
  evolution.
\newblock In {\em Proceedings of 12th ACM International Conference on Knowledge
  Discovery and Data Mining}, pages 44--54, 2006.

\bibitem{BZCLV07}
A.~Balasubramanian, Y.~Zhou, B.~Croft, B.N. Levine, and A.~Venkataramani.
\newblock Web search from a bus.
\newblock In {\em Proceedings of 2nd ACM Workshop on Challenged Networks (CHANTS)},
  pages 59--66, 2007.

\bibitem{Gephi}
M.~Bastian, S.~Heymann, and M.~Jacomy.
\newblock {Gephi: An open source software for exploring and manipulating
  networks}.
\newblock In {\em Proceedings of 3rd International AAAI Conference on Weblogs and Social
  Media}, 2009.

\bibitem{BCF09}
H.~Baumann, P.~Crescenzi, and P.~Fraigniaud.
\newblock {Parsimonious flooding in dynamic graphs}.
\newblock In {\em Proceedings of 28th ACM Symposium on Principles of Distributed
  Computing}, pages 260--269, 2009.

\bibitem{Berman96}
K.A. Berman.
\newblock {Vulnerability of scheduled networks and a generalization of Menger's
  Theorem}.
\newblock {\em Networks}, 28(3):125--134, 1996.

\bibitem{BF03}
S.~Bhadra and A.~Ferreira.
\newblock Complexity of connected components in evolving graphs and the
  computation of multicast trees in dynamic networks.
\newblock In {\em Proceedings of 2nd International Conference on Ad Hoc, Mobile and Wireless
  Networks (AdHoc-Now)}, pages 259--270, 2003.

\bibitem{BGLL08}
V.D. Blondel, J.L. Guillaume, R.~Lambiotte, and E.~Lefebvre.
\newblock {Fast unfolding of communities in large networks}.
\newblock {\em Journal of Statistical Mechanics: Theory and Experiment},
  2008:P10008, 2008.

\bibitem{BoP10}
C.~Boldrini and A.~Passarella.
\newblock { HCMM: Modelling spatial and temporal properties of human mobility
  driven by users' social relationships}.
\newblock {\em Computer Communications}, 33(9):1056--1074, 2010.

\bibitem{BSV07}
K.~Borne, S.~Sanyal, and A.Vespignani.
\newblock Network science.
\newblock {\em Annual Review of Information Science and Technology},
  41:537--607, 2007.

\bibitem{BFJ03}
B.~{Bui-Xuan}, A.~Ferreira, and A.~Jarry.
\newblock Computing shortest, fastest, and foremost journeys in dynamic
  networks.
\newblock {\em International Journal of Foundations of Comp. Science}, 14(2):267--285, April
  2003.

\bibitem{BGJL06}
J.~Burgess, B.~Gallagher, D.~Jensen, and B.N. Levine.
\newblock {Maxprop: Routing for vehicle-based disruption-tolerant networks}.
\newblock In {\em Proceedings of 25th IEEE Conference on Computer Communications
  (INFOCOM)}, pages 1--11, 2006.

\bibitem{CBD02}
T.~Camp, J.~Boleng, and V.~Davies.
\newblock Diameter of the world-wide web.
\newblock {\em Wireless Communications and Mobile Computing}, 2(5):483Ð 502,
  2002.

\bibitem{CLW07}
I.~Cardei, C.~Liu, and J.~Wu.
\newblock Routing in Wireless Networks with Intermittent Connectivity.
\newblock In {\em Encyclopedia of Wireless and Mobile Communications}, CRC
  Press, Taylor \& Francis, 2007.

\bibitem{CasCF09}
A.~Casteigts, S.~Chaumette, and A.~Ferreira.
\newblock Characterizing topological assumptions of distributed algorithms in
  dynamic networks.
\newblock In {\em Proceedings of 16th International Colloquium on Structural Information and Communication Complexity (SIROCCO)}, pages 126--140, 2009. (Full version in {\it arXiv:1102.5529}.) 

\bibitem{CasFMS10}
A.~Casteigts, P.~Flocchini, B.~Mans, and N.~Santoro.
\newblock Deterministic computations in time-varying graphs: Broadcasting under
  unstructured mobility.
\newblock In {\em Proceedings of 5th IFIP Conference on Theoretical Computer Science
  (TCS)}, pages 111--124, 2010.

\bibitem{CasFMS11}
A.~Casteigts, P.~Flocchini, B.~Mans, and N.~Santoro.
\newblock Measuring temporal lags in delay-tolerant networks.
\newblock In {\em Proceedings of 25th IEEE International Parallel and Distributed
  Processing Symposium (IPDPS)}, 2011.

\bibitem{ChHCDGS07}
A.~Chaintreau, P.~Hui, J.~Crowcroft, C.~Diot, R.~Gass, and J.~Scott.
\newblock Impact of human mobility on opportunistic forwarding algorithms.
\newblock {\em IEEE Transactions on Mobile Computing}, 6(6):606--620, 2007.

\bibitem{ChMMD08}
A.~Chaintreau, A.~Mtibaa, L.~Massoulie, and C.~Diot.
\newblock The diameter of opportunistic mobile networks.
\newblock {\em Communications Surveys \& Tutorials}, 10(3):74--88, 2008.

\bibitem{CleMMPS08}
A.~Clementi, C.~Macci, A.~Monti, F.~Pasquale, and R.~Silvestri.
\newblock {Flooding time in edge-markovian dynamic graphs}.
\newblock In {\em Proceedings of 27th ACM Symposium on Principles of Distributed Computing (PODC)}, pages 213--222, 2008.

\bibitem{CleP10}
A.~Clementi and F.~Pasquale.
\newblock {Information spreading in dynamic networks: An analytical approach}.
\newblock In: S. Nikoletseas, and J. Rolim (Eds), {\em Theoretical Aspects of
  Distributed Computing in Sensor Networks}, Springer, 2010.

\bibitem{Graphstream}
A.~Dutot, F.~Guinand, D.~Olivier, and Y.~Pign{\'e}.
\newblock {Graphstream: A tool for bridging the gap between complex systems and
  dynamic graphs}.
\newblock In {\em Proceedings of of Emergent Properties in Natural and Artificial Complex Systems. (EPNACS)}, pages 63--72, 2007.

\bibitem{Eagle06}
N.~Eagle and A.~(Sandy)~Pentland.
\newblock Reality mining: sensing complex social systems.
\newblock {\em Personal Ubiquitous Comput.}, 10(4):255--268, 2006.

\bibitem{EBM08}
J.~Eriksson, H.~Balakrishnan, and S.~Madden.
\newblock Cabernet: Vehicular content delivery using {W}i{F}i.
\newblock In {\em Proceedings of 14th ACM/IEEE International Conference on Mobile Computing and
  Networking}, pages 199--210, 2008.

\bibitem{Fer04}
A.~Ferreira.
\newblock Building a reference combinatorial model for {MANETs}.
\newblock {\em IEEE Network}, 18(5):24--29, 2004.

\bibitem{FKMS12}
P.~Flocchini, M.~Kellett, P.~Mason, and N.~Santoro.
\newblock Searching for black holes in subways.
\newblock {\em Theory of Computing Systems},
  50(1):158--184, 2012.

\bibitem{FMS09}
P.~Flocchini, B.~Mans, and N.~Santoro.
\newblock Exploration of periodically varying graphs.
\newblock In {\em Proceedings of 20th International Symposium on Algorithms and Computation
  (ISAAC)}, pages 534--543, 2009.

\bibitem{FCF11}
A~{F}riggeri, G~{C}helius, and E~{F}leury.
\newblock {E}go-munities, {E}xploring {S}ocially {C}ohesive {P}erson-based
  {C}ommunities.
\newblock Technical Report 7535, INRIA, 2011.

\bibitem{GAS11}
F.~Greve, L.~Arantes, and P.~Sens.
\newblock What model and what conditions to implement unreliable failure
  detectors in dynamic networks?
\newblock In {\em 3rd Workshop on Theoretical Aspects of Dynamic Distributed
  Systems (TADDS)}, 2011.

\bibitem{GP10}
P.~Grindrod and M.~Parsons.
\newblock {Social networks: Evolving graphs with memory dependent edges}.
\newblock Technical report, MPS\_2010-02, University of Reading, 2010.

\bibitem{GroV03}
M.~Grossglauser and M.~Vetterli.
\newblock {Locating nodes with EASE: Last encounter routing in ad hoc networks
  through mobility diffusion}.
\newblock In {\em Proceedings of 22nd Conference on Computer Communications (INFOCOM)},
  volume~3, pages 1954--1964, 2003.

\bibitem{GK07}
S.~Guo and S.~Keshav.
\newblock {Fair and efficient scheduling in data ferrying networks}.
\newblock In {\em Proceedings of ACM Conference on Emerging Network Experiment and
  Technology}, pages 1--13, 2007.

\bibitem{HG97}
F.~Harary and G.~Gupta.
\newblock {Dynamic graph models}.
\newblock {\em Mathematical and Computer Modelling}, 25(7):79--88, 1997.

\bibitem{Holme05}
P.~Holme.
\newblock {Network reachability of real-world contact sequences}.
\newblock {\em Physical Review E}, 71(4):46119, 2005.

\bibitem{HSPH07}
W.~Hsu, T.~Spyropoulos, K.~Psounis, and A.~Helmy.
\newblock {Modeling time-variant user mobility in wireless mobile networks}.
\newblock In {\em Proceedings of 26th IEEE Conference on Computer Communications
  (INFOCOM)}, pages 758--766, 2007.

\bibitem{IW11}
D.~Ilcinkas and A.M.~Wade.
\newblock On the power of waiting when exploring public transportation systems.
\newblock In {\em Proceedings of 15th International Conference on Principles of Distributed Systems (OPODIS)}, pages 451--464, 2011.

\bibitem{JMR10}
P.~Jacquet, B.~Mans, and G.~Rodolakis.
\newblock {Information propagation speed in mobile and delay tolerant
  networks}.
\newblock {\em IEEE Transactions on Information Theory}, 56(10):5001--5015, 2009.

\bibitem{JFP04}
S.~Jain, K.~Fall, and R.~Patra.
\newblock {Routing in a delay tolerant network}.
\newblock In {\em Proceedings of Conference on Applications, Technologies, Architectures, and Protocols for Computer Communications (SIGCOMM)}, pages 145--158, 2004.

\bibitem{JLW07}
E.P.C. Jones, L.~Li, J.K. Schmidtke, and P.A.S. Ward.
\newblock {Practical routing in delay-tolerant networks}.
\newblock {\em IEEE Transactions on Mobile Computing}, 6(8):943--959, 2007.

\bibitem{KY05}
H.~Kakugawa and M.~Yamashita.
\newblock {A dynamic reconfiguration tolerant self-stabilizing token
  circulation algorithm in ad-hoc networks}.
\newblock {\em 9th International Conference on Principles of Distributed Systems (OPODIS)}, pages 256--266, 2005.

\bibitem{KeK02}
D.~Kempe and J.~Kleinberg.
\newblock Protocols and impossibility results for gossip-based communication
  mechanisms.
\newblock In {\em Proceedings of 43rd Symposium on Foundations of Computer Science
  (FOCS)}, pages 471--480, 2002.

\bibitem{KKK00}
D.~Kempe, J.~Kleinberg, and A.~Kumar.
\newblock {Connectivity and inference problems for temporal networks}.
\newblock In {\em Proceedings of 32nd ACM Symposium on Theory of Computing (STOC)}, page
  513, 2000.

\bibitem{kempe03}
D.~Kempe, J.~Kleinberg, and E.~Tardos.
\newblock Maximizing the spread of influence through a social network.
\newblock In {\em Proceedings of 9th ACM International Conference on Knowledge
  Discovery and Data Mining (KDD)}, pages 137--146, 2003.

\bibitem{KerO09}
A.~Ker{\"a}nen and J.~Ott.
\newblock {DTN over aerial carriers}.
\newblock In {\em Proceedings of 4th ACM Workshop on Challenged Networks}, pages
  67--76, 2009.

\bibitem{KosKW08}
G.~Kossinets, J.~Kleinberg, and D.~Watts.
\newblock {The structure of information pathways in a social communication
  network}.
\newblock In {\em Proceedings of 14th ACM International Conference on Knowledge Discovery and Data Mining (KDD)}, pages 435--443, 2008.

\bibitem{Kostakos09}
V.~Kostakos.
\newblock {Temporal graphs}.
\newblock {\em Physica A}, 388(6):1007--1023, 2009.

\bibitem{KLO10}
F.~Kuhn, N.~Lynch, and R.~Oshman.
\newblock {Distributed computation in dynamic networks}.
\newblock In {\em Proceedings of 42nd ACM Symposium on Theory of Computing (STOC)}, pages
  513--522, 2010.

\bibitem{KMO11}
F.~Kuhn, and Y.~Moses, and R.~Oshman.
\newblock Coordinated consensus in dynamic networks.
\newblock In {\em 30th ACM symposium on Principles of Distributed Computing (PODC)}, pages 1--10, 2011.

\bibitem{LESK10}
J.~Leskovec, D.~Chakrabarti, J.~M. Kleinberg, C.~Faloutsos, and Z~Ghahramani.
\newblock Kronecker graphs: An approach to modeling networks.
\newblock {\em Journal of Machine Learning Research}, 11:985--1042, 2010.

\bibitem{LeKF07}
J.~Leskovec, J.~Kleinberg, and C.~Faloutsos.
\newblock Graph evolution: Densification and shrinking diameters.
\newblock {\em ACM Transactions on Knowledge Discovery from Data}, 1(1), 2007.

\bibitem{LDS03}
A.~Lindgren, A.~Doria, and O.~Schel\'{e}n.
\newblock Probabilistic routing in intermittently connected networks.
\newblock {\em Mobile Computing and Communications Review}, 7(3):19--20, 2003.

\bibitem{LMS99}
I.~Litovsky, Y.~M\'etivier, and E.~Sopena.
\newblock {\em Graph relabelling systems and distributed algorithms}.
\newblock In: H. Ehrig, H.J. Kreowski, U. Montanari and G. Rozenberg (Eds.),
  {\em Handbook of Graph Grammars and Computing by Graph Transformation}, World
  Scientific Publishing, 1999.

\bibitem{LW09b}
C.~Liu and J.~Wu.
\newblock Scalable routing in cyclic mobile networks.
\newblock {\em IEEE Transactions on Parallel and Distributed Systems}, 20(9):1325--1338, 2009.

\bibitem{MSG08}
  Y.~Mah{\'e}o, R.~Said, and F.~Guidec,
  \newblock Middleware support for delay-tolerant service provision in disconnected mobile ad hoc networks.
  \newblock In {\em Proceedings of 22nd IEEE Parallel and Distributed Processing Symposium (IPDPS)}, pages 1--6, 2008.


\bibitem{OW05}
R.~O'Dell and R.~Wattenhofer.
\newblock Information dissemination in highly dynamic graphs.
\newblock In {\em Proceedings of Joint Workshop on Foundations of Mobile Computing
  (DIALM-POMC)}, pages 104--110, 2005.

\bibitem{PS00}
M.~Papadopouli and H.~Schulzrinne.
\newblock {Seven degrees of separation in mobile ad hoc networks}.
\newblock In {\em Proceedings of IEEE Global Communication Conference (GLOBECOM)},
  volume~3, pages 1707--1711, 2000.

\bibitem{RDBT12}
P.~Ruiz, B.~Dorronsoro, P.~Bouvry, L,~Tard{\'o}n. 
\newblock Information dissemination in VANETs based upon a tree topology.
\newblock {\em Ad Hoc Networks}, 10(1):111--127, 2012.

\bibitem{RBK07}
R.~Ramanathan, P.~Basu, and R.~Krishnan.
\newblock Towards a formalism for routing in challenged networks.
\newblock In {\em Proceedings of 2nd ACM Workshop on Challenged Networks (CHANTS)},
  pages 3--10, 2007.

\bibitem{RRS11}
F.J. Ros, P.M. Ruiz, and I.~Stojmenovic.
\newblock Acknowledgment-based broadcast protocol for reliable and efficient
  data dissemination in vehicular ad-hoc networks.
\newblock {\em IEEE Transactions on Mobile Computing}, 11(1):33--46, 2012.

\bibitem{SQFCA11}
N.~Santoro, W.~Quattrociocchi, P.~Flocchini, A.~Casteigts, and F.~Amblard.
\newblock Time-varying graphs and social network analysis: Temporal indicators
  and metrics.
\newblock {\em 3rd AISB Social Networks and Multiagent Systems Symposium (SNAMAS)}, pages 32--38, 2011.

\bibitem{Scherrer08}
A.~Scherrer, P.~Borgnat, E.~Fleury, J.~L. Guillaume, and C.~Robardet.
\newblock Description and simulation of dynamic mobility networks.
\newblock {\em Computer Networks}, 52(15):2842--2858, 2008.

\bibitem{SPR05}
T.~Spyropoulos, K.~Psounis, and C.S. Raghavendra.
\newblock {Spray and wait: an efficient routing scheme for intermittently
  connected mobile networks}.
\newblock In {\em Proceedings of ACM Workshop on Delay-Tolerant Networking}, page 259,
  2005.

\bibitem{TMML10}
J.~Tang, M.~Musolesi, C.~Mascolo, and V.~Latora.
\newblock {Characterising temporal distance and reachability in mobile and
  online social networks}.
\newblock {\em ACM Computer Communication Review}, 40(1):118--124, 2010.

\bibitem{TMMLN10}
J.~Tang, M.~Musolesi, C.~Mascolo, V.~Latora, and V.~Nicosia.
\newblock {Analysing information flows and key mediators through temporal
  centrality metrics}.
\newblock In {\em Proceedings of the 3rd Workshop on Social Network Systems},
  pages 1--6. ACM, 2010.

\bibitem{TSM+09}
J.~Tang, S.~Scellato, M.~Musolesi, C.~Mascolo, and V.~Latora.
\newblock {Small-world behavior in time-varying graphs}.
\newblock {\em Physical Review E}, 81(5):55101, 2010.

\bibitem{WTSB09}
S.~Wang, J.L. Torgerson, J.~Schoolcraft, and Y.~Brenman.
\newblock The deep impact network experiment operations center monitor and
  control system.
\newblock In {\em Proceedings of 3rd IEEE international Conference on Space Mission Challenges For information Technology}, pages 34--40, 2009.

\bibitem{YamK96}
M.~Yamashita and T.~Kameda.
\newblock Computing on anonymous networks: Part {I} and {II}.
\newblock {\em IEEE Transactions on Parallel and Distributed Systems}, 7(1):69--96, 1996.

\bibitem{ZKL+07}
X.~Zhang, J.~Kurose, B.N. Levine, D.~Towsley, and H.~Zhang.
\newblock {Study of a bus-based disruption-tolerant network: mobility modeling
  and impact on routing}.
\newblock In {\em Proceedings of 13th ACM International Conference on Mobile Computing and Networking}, pages 195--206, 2007.

\bibitem{Zha06}
Z.~Zhang.
\newblock {Routing in intermittently connected mobile ad hoc networks and delay
  tolerant networks: Overview and challenges}.
\newblock {\em IEEE Communications Surveys \& Tutorials}, 8(1):24--37, 2006.

\end{thebibliography}
\end{document}